\documentclass[prx,prl,a4paper,aps,twocolumn,superscriptaddress,longbibliography]{revtex4-2}
\usepackage{amssymb,amsfonts,bm}
\usepackage{graphicx}
\usepackage{epstopdf}
\usepackage{times}
\usepackage{float}
\usepackage{lipsum}
\usepackage{color}
\usepackage{dcolumn}
\usepackage{bm}
\usepackage{subfigure}
\usepackage[colorlinks,linkcolor=blue,anchorcolor=blue,urlcolor=blue,citecolor=blue]{hyperref}
\usepackage{verbatim} 
\usepackage{amsmath} 
\usepackage{tikz}
\usepackage{braket}
\usepackage{bbold}

\usepackage{amsthm}

\def\kk{\rangle\!\rangle}

\def\map{\mathcal}

\newtheorem{Def}{Definition}
\newtheorem{thm}{\protect\theoremname}

\newtheorem{lem}{Lemma}

\providecommand{\theoremname}{Theorem}
\newcommand*{\id}{\mathbb{1}}
\newcommand*{\Tr}{\textrm{Tr}}

\usepackage{hyperref}
\hypersetup{colorlinks=true, linkcolor=blue, citecolor=red, urlcolor=blue  }

\usepackage{physics}

\begin{document}

\title{Inferring the arrow of time in quantum spatiotemporal correlations }

\author{Xiangjing Liu} 
\email{liuxj@mail.bnu.edu.cn}
\affiliation{Department of Physics, Southern University of Science and Technology, Shenzhen 518055, China}
\affiliation{Department of Physics, City University of Hong Kong, 83 Tat Chee Avenue, Kowloon, Hong Kong SAR, China}

\author{Qian Chen}
\email{chenqian.phys@gmail.com}
\affiliation{Univ Lyon, Inria, ENS Lyon, UCBL, LIP, F-69342, Lyon Cedex 07, France}

\author{Oscar Dahlsten}
\email{oscar.dahlsten@cityu.edu.hk}
\affiliation{Department of Physics, City University of Hong Kong, 83 Tat Chee Avenue, Kowloon, Hong Kong SAR, China}
\affiliation{Shenzhen Institute for Quantum Science and Engineering, Southern University of Science and Technology, Shenzhen 518055, China}
\affiliation{Institute of Nanoscience and Applications, Southern University of Science and Technology, Shenzhen 518055, China}



\begin{abstract}  
We consider how to tell the time-ordering associated with measurement data from quantum experiments at two times and any number of qubits. We define an arrow of time inference problem. We consider conditions on the initial and final states that are symmetric or asymmetric under time reversal. We represent the spatiotemporal measurement data via the pseudo density matrix space-time state. There is a forward process which is CPTP and a reverse process which is obtained via a novel recovery map based on inverting unitary dilations. For asymmetric conditions, the protocol determines whether the data is consistent with the unitary dilation recovery map or the CPTP map. For symmetric conditions, the recovery map yields a valid CPTP map and the experiment may have taken place in either direction. We also discuss adapting the approach to the Leifer-Spekkens or Process matrix space-time states. 
\end{abstract}

\maketitle


\section{Introduction}
The arrow of time refers to the apparent asymmetry between time moving forward and backward. One can often tell if a movie is being played in the correct, forward, direction or not. Understanding the arrow of time and time-reversal symmetry is of long-running foundational interest, whether in cosmology, particle physics or thermodynamics~\cite{halliwell1996physical,hawking1985arrow,prigogine1978time,sozzi2008discrete,carroll2010eternity,seif2021machine}.

The recent renewed interest in creating a unified spatiotemporal framework for quantum theory~\cite{leggett1985quantum,hardy2007towards,chiribella2009theoretical,oreshkov2012quantum,fitzsimons2015quantum,leifer2008quantum,jia2023spatiotemporal,huang2022leggett,song2023causal}, wherein space and time are treated on a more equal footing, raises new questions and challenges concerning the arrow of time. Temporal correlations are being analyzed with tools originally created for spatial correlations~\cite{bell1964einstein,bennett1993teleporting,dakic2012quantum,rio2011thermodynamic,nielsen2010quantum}. Leggett and Garg demonstrated that quantum systems display a type of timelike correlation unexplainable by macroscopic realism~\cite{leggett1985quantum}.  Techniques have been developed to certify quantum temporal correlation in quantum information theory~\cite{shrotriya2022certifying,chen2023semi,chiribella2022nonequilibrium,brukner2004quantum,ku2018hierarchy,PhysRevA.107.040101}. Genuine temporal signals have been utilized to infer quantum causal structure ~\cite{fitzsimons2015quantum,liu2023quantum}. People also found that causal structures in quantum theory can be superposed~\cite{hardy2007towards,chiribella2009theoretical,oreshkov2012quantum} with consequences for information communication capacity and thermodynamics~\cite{Ebler2018enhanced,nie2022exp,felce2020quantum,cao2022quantum,liu2022thermodynamics}. Of particular interest here is that in analyzing temporal correlations with tools for spatial correlations an important new feature emerges: the potential time-asymmetry of temporal correlations~\cite{di2021arrow,liu2023unification,marletto2019theoretical,hawking1985arrow,bai2022quantum}. This gives rise to the question of how this time asymmetry manifests itself in the spatiotemporal data, and the closely related question of how one can determine the temporal ordering given the data.
\begin{figure}
  \centering
  \includegraphics[scale=0.24]{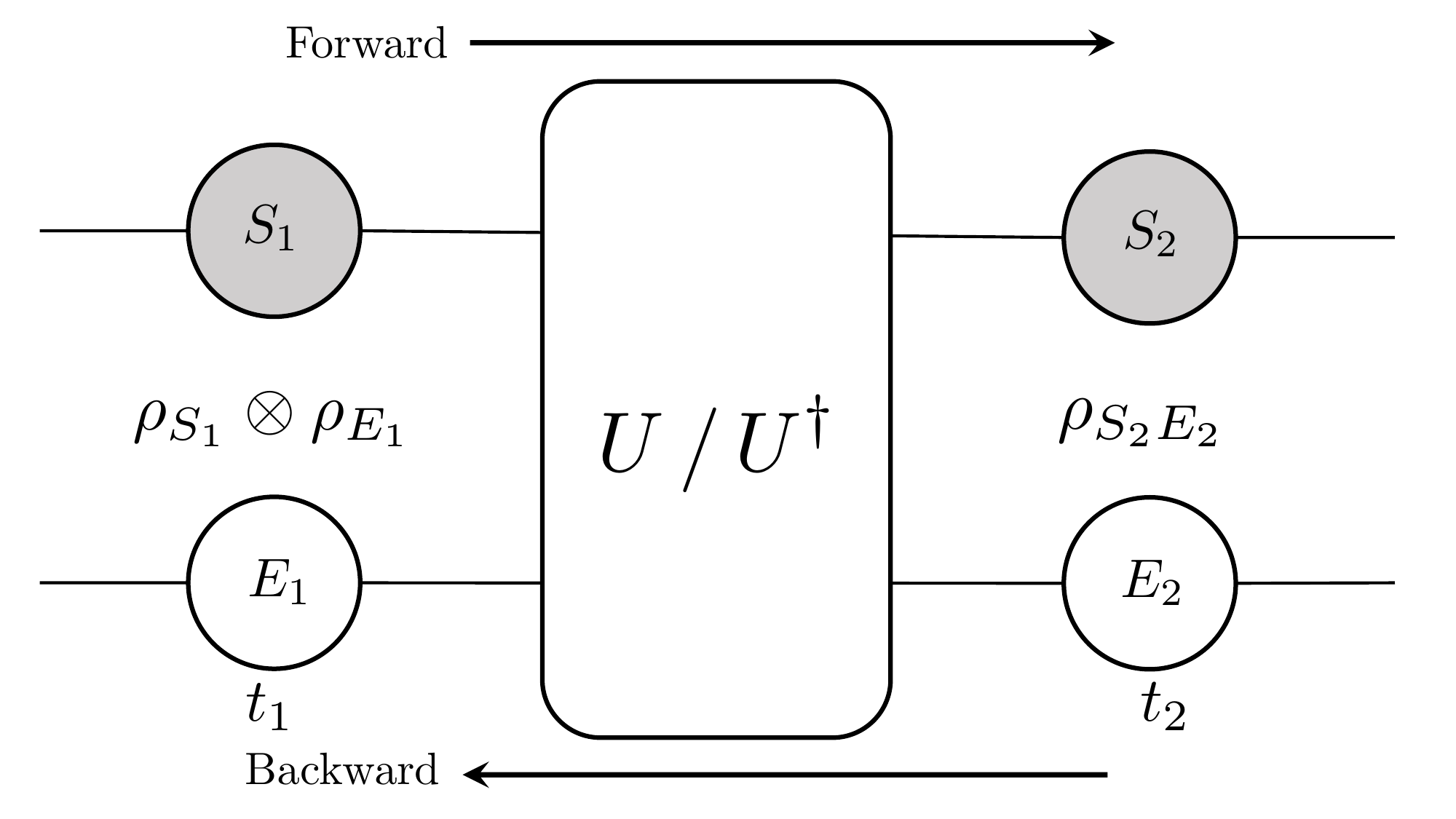}
  \caption{The task is to determine whether a forward process generated the data or whether the time labels of the data have been switched. 
   {\bf{ Forward Process.}}  
    A crucial initial condition is that the system $S$ and the environment $E$ start in a product state $\rho_{S_1} \otimes \rho_{E_1}$ at time $t_1$.  Undergoing a unitary interaction $U$, the joint system $SE$ becomes in general correlated at time $t_2$. Ignoring $E$, the dynamics of the system $S$ are characterized by a CPTP map.  
      {\bf{ Backward Process.}}  At $t_2$, the system $S$ with the environment $E$ starts in a correlated state $\rho_{S_2E_2} $. The joint system $SE$ then undergoes the backward evolution $U^\dag$. In this scenario, the dynamics of the system $S$ is given by a recovery map that is positive but not necessarily  CPTP.
 }
  \label{fig:setup}
\end{figure}

In this work, we accordingly give a protocol determining the arrow of time given quantum spatiotemporal correlations. We employ the so-called pseudo-density matrix (PDM) formalism~\cite{fitzsimons2015quantum}. The PDM is constructed operationally, from measurements at several locations and times. There is, as depicted in Fig.~\ref{fig:setup}, a forward process, which is a completely positive and trace-preserving (CPTP) map of the system state. Flipping the time label of the collected data gives rise to an associated reverse process, a recovery map of independent interest, that turns out to be equivalent to inverting the unitary dilation of the forward channel.  When the conditions on the initial and final states are asymmetric, our protocol helps determine whether the data aligns with the unitary dilation recovery map or the CPTP map. In symmetric conditions, the recovery map produces a valid CPTP map, indicating that the experiment could have taken place in either direction.  The main technical contribution is a method for extracting a matrix representing the dynamics (the Choi Jamiolkowski operator) given the spatio-temporal data, via a vectorisation of the PDM.

\section{Arrow of time inference problem}
 We first briefly review the PDM formalism used for formulating the problem. The PDM generalizes the density matrix by assigning a Hilbert space to each instant in time. In particular, the two-time PDM $R_{12} \in \map B (H_1 \otimes H_2)$, which aligns with the arrow of time and will be used in this work, is defined as~\cite{fitzsimons2015quantum,liu2023quantum}
\begin{align}\label{eq: DefForwardPDM}
\text{ (forward)} \,& R_{12} = \frac{1}{4^{n}} \sum^{4^n-1}_{i_1,i_2 =0 }  \langle \{ \sigma_{i_1}, \sigma_{i_2} \} \rangle  \sigma_{i_1} \otimes  \sigma_{i_2} , 
\end{align}
where  $ \sigma_{i_\alpha} \in \{ \id, \sigma_x, \sigma_y, \sigma_z \} ^{\otimes n} $ is an $n$-qubit Pauli matrix at time $t_\alpha$. $\sigma_{i_\alpha}$ is extended to an observable associated with two times, $ \sigma_{i_1} \otimes  \sigma_{i_2}$ that has expectation value  $\langle \{ \sigma_{i_1}, \sigma_{i_2} \} \rangle $. One measurement scheme for obtaining   $\langle \{ \sigma_{i_1}, \sigma_{i_2} \} \rangle $ is provided in Sec.~\ref{sec:dynamics}. We refer to Eq.~\eqref{eq: DefForwardPDM} as the forward PDM. The partial trace of a PDM still gives a valid PDM~\cite{liu2023quantum}. The PDM is Hermitian with unit trace but may have negative eigenvalues. The negative eigenvalue is a sufficient but not necessary condition for quantum temporal correlation, as the density matrix at a single time cannot explain it. 

While negative eigenvalues signify quantum temporal correlation, the arrow of time remains a separate question. Quantum temporal correlation has a unique feature of being time asymmetric compared to quantum spatial correlations, such as quantum entanglement. The negative values in the PDM can be used to quantify the strength of quantum temporal correlation~\cite{fitzsimons2015quantum, liu2023quantum}, but they do not carry the information of the time asymmetry. This motivates the investigation of the arrow of time in the PDM formalism. 

The arrow of time inference problem is formulated in the PDM formalism as follows. Consider that an $n$-qubit system undergoes a CPTP map $\map E: \map B (H_1) \rightarrow \map B (H_2)$. In order to construct the 2-time PDM, an experimenter Alice implements the observables $\sigma_{i_1} $ at initial time $t_1$ and $ \sigma_{i_2}$ at final time $t_2$. She then collects the data $\{\langle \{ \sigma_{i_1}, \sigma_{i_2} \} \rangle \}  $ but forgets to record the time order of events. This means that, due to the lack of time information, that data could be used to construct two possible PDMs, one is the forward PDM listed in Eq.~\eqref{eq: DefForwardPDM} and the other is in the following
\begin{align}
\label{eq:DefbackwardPDM}
\text{(backward)} \,& \bar{R}_{12}= \frac{1}{4^{n}} \sum^{4^n-1}_{i_1, i_2=0}  \langle \{ \sigma_{i_1}, \sigma_{i_{2}} \} \rangle \sigma_{i_2} \otimes \sigma_{i_{1} } \,  .
\end{align}
We refer this to as the backward PDM.
Alice wants to know which one is consistent with the flow of time.

Let us first understand the problem better before we dive into it. The collected data $\{\langle \{ \sigma_{i_1}, \sigma_{i_2} \} \rangle \}  $ includes $ \{ \langle \{ \sigma_{i_1}, \id \} \rangle \}  $ and   $\{\langle \{   \id, \sigma_{i_2} \} \rangle \}  $,  i.e., scenarios where Alice does nothing at one time while measuring $n$-qubit Pauli matrices at another time.  Attaching those two pieces of data to the corresponding Pauli matrices and summing up gives two valid density matrices (up to a normalization constant).  Denote the two density matrices by
\begin{align}
&\rho := \frac{1}{2^n} \sum^{4^n-1}_{i_1=0}  \langle \{ \sigma_{i_1}, \id \} \rangle \sigma_{i_1} ,
 \gamma  := \frac{1}{2^n} \sum^{4^n-1}_{i_2=0}  \langle \{ \id, \sigma_{i_2} \} \rangle \sigma_{i_2}.\nonumber
\end{align}
It can be directly verified that 
\begin{align}
&\rho  =\Tr_2 R_{12} = \Tr_1 {\bar R}_{12}, \nonumber\\
&\gamma = \Tr_1 R_{12} = \Tr_2 {\bar R}_{12} .
\end{align}
To put it in words, the initial and final states in the two PDMs are swapped. This reveals a useful relation between the two PDMs
\begin{align} \label{eq: FBrelation}
\bar{R}_{12} = S \, R_{12}  \, S^\dag,
\end{align}
where $S:=\sum^{2^n-1}_{i,j=0} \ketbra{ij}{ji}= \frac{1}{2^n} \sum_i \sigma_i \otimes \sigma_i $ denotes the $n$-qubit swap operator. The swap operator here can be treated as a time reversal operation on a PDM.  It is natural to call the PDM $\bar R$ a time-reversed version of $R$. Therefore, Alice's task then becomes to distinguish whether her data table corresponds to the actual forwards process $R$ or $\bar R$. The idea of an arrow of time is associated with there being examples where $\bar R$ is not realizable but $R$ is, an asymmetry that will next be associated with the boundary conditions. 

\section{Boundary conditions for inferring arrow}
\label{sec: ArrowTime}
A common explanation for time asymmetry concerns the conditions on the initial and final states, i.e., the boundary conditions~\cite{hawking1985arrow}. Next, we will discuss inferring the arrow of time under both symmetric and asymmetric entropic boundary conditions.

\subsection{Asymmetric boundary conditions}
The asymmetry of boundary conditions often refers to the asymmetry of entropies. In quantum information processing as illustrated in the forward direction of FIG.~\ref{fig:setup}, the system of interest $S$ is often assumed to be initially in a product state with the environment $E$ at initial time $t_1$, i.e., $\rho_{S_1} \otimes \rho_{E_1}$. System $S$ then interacts with the environment $E$ via the unitary $U$, arriving at a final state $\rho_{S_2E_2}:= U(\rho_{S_1} \otimes \rho_{E_1} )U^\dag$. Let $\rho_{S_2}:=\text{Tr}_E \, \rho_{S_2E_2} $, $\rho_{E_2}:=\text{Tr}_S \rho_{S_2E_2} $. The sum of the entropies is non-decreasing during the process, i.e.,
\begin{align}
    S(\rho_{S_1}) +  S(\rho_{E_1}) \leq     S(\rho_{S_2}) +  S(\rho_{E_2})
\end{align}
where $S(\rho)$ denotes the von Neumann entropy and the subadditivity of entropy~\cite{nielsen2010quantum} is used.
The consequences of the inequality can be employed to infer the arrow of time.

For the forward evolution as shown in Fig.~\ref{fig:setup}, by ignoring the environment $E$, the open dynamics of the system $S$  can be characterized by a CPTP map  $\map E$, its action is given by 
\begin{align}
    \map E (\rho_{S_1}) :=   \text{Tr}_E \, U(\rho_{S_1} \otimes \rho_{E_1} )U^\dag.
\end{align}
Let $M\in \map B (H_1 \otimes H_2)$ denotes the Choi–Jamio{\l}kowski (CJ) matrix of $\map E$ and it is defined as~\cite{choi1975completely, jamiolkowski1972linear}
\begin{align}\label{eq:CJ}
M  &= \sum^{2^n-1}_{i,j=0} \ketbra{i}{j} \otimes \map E (\ketbra{j}{i}). 
\end{align}
The map $\map E$ being CP is equivalent to its Choi matrix $M^{T_1}$ being positive semidefinite, where $T_1$ denotes the transpose on $ \map B (H_1) $.

As for the backward evolution, the joint state $\rho_{S_2E_2}$ of the system-environment is, in general, correlated.
The unitary evolution $U^\dag$ undoes the correlations between $S$ and $E$. In this scenario, the local open dynamics $\bar{\map E}$ of the system $S$ is linear but, in general, cannot be characterized by a CP map~\cite{cao2023quantum,rivas2014quantum,breuer2016nonmarkovian,wei2023realizing}, i.e., its Choi matrix $\bar{M}^{T_1}$ is not positive semidefinite.

Based on the analysis above, the positivity of Choi matrices $M^{T_1}$ and $\bar{M}^{T_1}$ can be utilized to infer the arrow of time. If $M$ is positive semidefinite and $\bar M$ has negative eigenvalues, then the PDM $R$ is the one consistent with the arrow of time. However, in situations when both $M$ and $\bar M$ are positive semidefinite, we need more information to determine the arrow of time.

\subsection{Symmetric boundary conditions}

The entropic boundary conditions are symmetric when $S(\rho_{S_1})+S(\rho_{E_1}) = S(\rho_{S_2})+S(\rho_{E_2})$. One particular scenario is that the system undergoes a unitary evolution $\tilde U$. The forward dynamics of system $S$ is then given by $\map E$ with its action $\map E (\rho_{S_1}):= \tilde{U} \rho_{S_1} \tilde{U}^\dag $, where $\tilde{U}$ is part of $U$ which acts on the system alone. Naturally, the backward dynamics of $S$ is given by $\bar{ \map E} $ with the action $\bar{ \map E} (\rho_{S_2}):= \tilde{U}^\dag \rho_{S_2} \tilde{U} $. 
Therefore both the forward and backward maps are CPTP. 
Moreover, there are also interesting cases in quantum thermodynamics in which the backward map is also CPTP~\cite{crooks2008quantum}. Let $\tau$ denote the Gibbs state. When the system $S$ and the environment $E$ are fully thermalized, i,e., $U(\tau_{S_1} \otimes \tau_{E_1}) U^\dag= \tau_{S_2} \otimes \tau_{E_2}$, the backward local dynamics $\bar {\map E}$ on $S$, which is given by its action
\begin{align}
\bar{\map E}(\tau_{S_2})=\Tr_E\left( U^{\dagger}\tau_{S_2} \otimes\tau_{E_2} U\right), \nonumber
\end{align}
is a CPTP map. 
We thus need more information to infer the arrow of time, which we will discuss more on this point in section~\ref{sec:beyongPDM}.

\section{Extracting dynamics from spatiotemporal correlations}
\label{sec:dynamics}

 In this section, we show how to extract information about processes from the forward and backward PDMs.

Let us first introduce a closed-form of the PDM consisting of multiple qubits across two times~\cite{liu2023quantum}. In order to obtain a closed-form for the forward PDM, the measurement scheme for determining the expectation values $  \langle \{ \sigma_{i_1}, \sigma_{i_2} \} \rangle$ is crucial. 
If the measurement scheme for the observable $\sigma_i$ at each time is set to be the projectors that project the state onto the $\pm 1$ eigenspaces of $\sigma_i$, i.e., 
$$
\left\{\Pi^i_+ = \frac{\id + \sigma_i}{2} , \Pi^i_- = \frac{\id - \sigma_i}{2}  \right\},
$$
the expectation value  $  \langle \{ \sigma_{i_1}, \sigma_{i_2} \} \rangle$ of the product of $\sigma_{i_1}$ made at $t_1$ and  $\sigma_{i_2}$ made at $t_2$ is read as
\begin{align}
      \langle \{ \sigma_{i_1}, \sigma_{i_2} \} \rangle =&  \langle \{ \Pi^{i_1}_+, \Pi^{i_2}_+ \} \rangle +  \langle \{ \Pi^{i_1}_-, \Pi^{i_2}_- \} \rangle \nonumber\\
      &-\left(  \langle \{ \Pi^{i_1}_-, \Pi^{i_2}_+ \} \rangle +  \langle \{ \Pi^{i_1}_+, \Pi^{i_2}_- \} \rangle \right). \nonumber
\end{align}

Given this measurement scheme $\{ \Pi^i_+, \Pi^i_-\}$, the corresponding closed-form of a 2-time PDM is expressed by~\cite{liu2023quantum}
\begin{align}
\label{eq: forwardPDM}
R= \frac{1}{2} \left( (\rho\otimes \id_2) \, M + M \, (\rho \otimes \id_2) \right).
\end{align}
This closed-form expression has been taken as a definition for a quantum spatiotemporal framework, called symmetric bloom~\cite{fullwood2022quantum,parzygnat2023time,fullwood2023quantum}.
Similarly, given the coarse-grained measurement scheme, we define that the closed form of the backward PDM is given by 
\begin{align}
\label{eq: backwardPDM}
\bar R= \frac{1}{2} \left( (\gamma \otimes \id_1) \, \bar M + \bar M \, (\gamma \otimes \id_1) \right),
\end{align}
where $\bar M$ denotes the CJ matrix of the backward process $\bar {\map E}$. A key justification for defining Eq.~\eqref{eq: backwardPDM} is the case of unitary evolutions, as will be described around Eq.~\eqref{eq:backwardU1} below.

The map $\bar{\map E}$, which is of independent interest, can be defined as follows. 
\begin{Def}[Unitary dilation recovery map]
Consider a CPTP map $\map E$ with input space $S_1$ and output space $S_2$ and a unitary dilation acting on $SE$ to give the output state $\rho_{S_2E_2}$ (we again denote the output $\rho_{S_2}$ as $\gamma$) . For a given valid unitary dilation $U$ and initial state $\rho_{S_1} \otimes \rho_{E_1}$, we can define a  unique recovery map $\bar {\map E}$ as the map corresponding to a CJ matrix $\bar M$ which respects
\begin{equation}\label{eq: RecoveryMap}
\Tr_{E_1E_2}\bar{R}_{S_1E_1S_2E_2}:= \frac{1}{2} \left( (\gamma \otimes \id_1) \, \bar M + \bar M \, (\gamma \otimes \id_1) \right),
\end{equation}
where $\bar{R}_{S_1E_1S_2E_2}$ is the inverse PDM of the process, associated with taking $\rho_{S_2E_2}$  as the initial state and then applying the inverse global evolution $U^\dag$ . We will show in Sec.\ref{sec:backward process} how $\bar M$ can be extracted from Eq.\eqref{eq: RecoveryMap}.
\end{Def}

The map $\bar{\map E}$ defined by Eq.~\eqref{eq: RecoveryMap} is in the form of its CJ representation $\bar M$, i.e.,
\begin{align}
    \bar{M}  = \sum^{2^n-1}_{i,j=0} \ketbra{i}{j} \otimes \bar{ \map E} (\ketbra{j}{i}).
    \end{align}
Therefore, given the input state $\rho_{S_1E_1}$ and the unitary evolution $U$, the operation $\bar{\map E}$ can be solved from Eq.~\eqref{eq: RecoveryMap}. 
Moreover, the output of the map $\bar{\map E}$ for any input $\rho$  can be calculated in the CJ representation via  
\begin{align}
    \bar{\map E}  (\rho) = \Tr_1 (\rho \otimes \id_2  )\bar M .
\end{align}
The operation $\bar{\map E}$ on the system $S$ is linear but may not be a CP map. However, we note that if the final state $\rho_{S_2E_2}$ is in a product state, then $\bar M$ must represent a CP map.

The justification for the closed form of the backward PDM $\bar R$  and the unitary dilation recovery map $\bar{\map E}$, for the case of unitary evolution $\map E(\cdot) = \tilde{U}(\cdot) \tilde{U}^\dag$, is below. Recall the relation between the forward and backward PDMs in Eq.~\eqref{eq: FBrelation}, direct calculation and some manipulation show that
\begin{align}\label{eq:backwardU0}
\bar{R} =& S \, R \, S^{\dag}= \frac{1}{2}\sum_{i,j}   \map E (\ketbra{i}{j})  \otimes( \rho\ketbra{j}{i} +\ketbra{j}{i} \rho)  \nonumber\\
=& \frac{1}{2}\sum_{i,j}   \map E(  \rho\ketbra{j}{i} +\ketbra{j}{i} \rho )\otimes \ketbra{i}{j} ,
\end{align}
where the linearity of the forward map $\map E$ is used. Given that $ \map E(\cdot) = \tilde{U}(\cdot) \tilde{U}^\dag$, then $\map E(\rho \ketbra{i}{j})= \map E(\rho) \map E( \ketbra{i}{j}) $. The reversed PDM $\bar R$ can be written as
\begin{align}\label{eq:backwardU1}
\bar R &= \frac{1}{2}\sum_{i,j} \left(  \map E (  \rho) \map E(\ketbra{j}{i}) + \map E (\ketbra{j}{i})  \map E( \rho) )\otimes \ketbra{i}{j}   \right) \nonumber\\
&= \frac{1}{2}\sum_{i,j} \left(  \map E (  \rho) \ketbra{j}{i} + \ketbra{j}{i}  \map E( \rho) )\otimes {\map E}^\dag (\ketbra{i}{j} )  \right) \nonumber\\
 &= \frac{1 }{2} (\bar M \, \map E(\rho) + \map E (\rho)  \,\bar M )
\end{align}
where $ {\map E}^\dag$ is the Hilbert-Schmidt adjoint of $\map E$ and $ \sum_{ij} \ketbra{i}{j} \otimes \map E (\ketbra{j}{i})  = \sum_{ij}  {\map E}^\dag(\ketbra{i}{j}) \otimes \ketbra{j}{i} $ is used in the second equality. Therefore, $\bar M= \sum_{i,j} \ketbra{i}{j} \otimes \tilde{U}^\dag (\ketbra{j}{i}) \tilde{U}$.
In this case, the Choi matrices $M^{T_1}, \bar{M}^{T_1}$ are positive.

From the closed-form expressions in Eqs.~\eqref{eq: forwardPDM} and~\eqref{eq: backwardPDM}, one can directly observe that the information about the processes, namely the CJ matrices $M$ and $\bar M$, is encoded in $R$ and $\bar R$, respectively. Next, we demonstrate how to extract $M$ and $\bar M$ from the forward and backward PDMs, respectively.

\subsection{Forward process}

Let $\map R: \map B(H_1 \otimes H_2) \rightarrow \map B (H_1 \otimes H_2) $ be the linear operator, defined by 
\begin{align}
\map R_{\rho} (M )=&   \frac{1}{2} ( \rho \, M +M \, \rho ).
\end{align}
Then, the extraction of $M$ can be treated as finding the inverse map of $\map{R }_{\rho}$. Note that, for the sake of simplicity in notation, the identity operator $\id_2 \in \map B (H_2)$ is omitted. In the following, one should tensor product identity operators in suitable bounded operator spaces when there is a mismatch in dimensionality.

As the standard treatment in quantum information, our method for finding the inverse map utilizes the vectorization of operators. Given a generic quantum operator 
$ O= \sum_{ij}\map O_{ij} \ket{i}\bra{j} \in \mathcal{B}(H)  $, its vectorization is expressed by 
$$
 | O \kk = \sum_{ij} O_{ij} \ket{i} \otimes \ket{j}  \in H \otimes H.
$$ 
The vectorization of $\map R_{\rho} (M)(=R) $ is then given by
\begin{align} \label{eq:VectorForwardPDM}
| \map {R}_{\rho} ( M)  \kk =& \frac{1}{2} | \rho \, M+ M \, \rho  \kk   \nonumber\\
=& \frac{1}{2} \left( \rho \otimes \mathbb{1} +  \mathbb{1} \otimes \rho^T \right)  | M \kk  \nonumber\\
=:& A |M \kk,
\end{align}
where $ | BCD \kk=D \otimes B^T \, |C \kk $ with $B,C,D \in  \map B(H_1\otimes H_2)$  is used in the second equality~\cite{wood2011tensor} and $\id  $ denotes the identity operator in $\map B (H_1 \otimes H_2)$.
Therefore, the invertibility of $\mathcal{R}_{\rho} $  resorts to the invertibility of the operator $A$, and thus the invertibility of $\rho$. 

If $\rho$ is full rank, then the operator $A$ and the map $\map {R}_{\rho}$ are invertible.  
 The inverse map $\mathcal {R}_{\rho}^{-1}$ can be defined via
\begin{align}
| \mathcal{R}_{\rho}^{-1} (O)  \kk = A^{-1} | O \kk  ,
\end{align}
where $O \in {\map B}(H_1 \otimes H_2)$.
Therefore, one has 
\begin{align}
| \mathcal{R}_{\rho}^{-1} (R )  \kk= A^{-1} A | M \kk = | M \kk  .
\end{align}
where Eq.~\eqref{eq:VectorForwardPDM} is used in the first equality.
Thus, one has obtained the full information of CJ matrix $M$ of the forward process.
Alternatively, one can also employ the following Lemma to get the analytic expression of $M$.
\begin{lem} 
\label{lem: SylvesterEq}
 Let $A$ and $B$ be operators whose spectra are contained in the open right half-plane and the open left half-plane, respectively. Then the solution of the operator equation 
 \begin{align} \label{eq: Sylvester}
 A X - X B = Y,
 \end{align}
  can be expressed as ~\cite{bhatia1997and}
 \begin{align}
 X= \int_0^\infty e^{-tA} \, Y \, e^{tB} dt .
 \end{align}
\end{lem}
\noindent The closed-form of PDM  is of the form of Eq.\eqref{eq: Sylvester}. 
 Given that $\rho $ is full rank, according to Lemma~\ref{lem: SylvesterEq}, the expression for $M$ is given by 
\begin{align}
M=2 \int_0^\infty e^{-t \rho} \, R \, e^{-t \rho} dt  .
\end{align}

If $\rho$ is rank deficient, $A$ is also rank deficient therefore $\map {R}_{\rho}$ is not invertible. However, some information about the quantum process $M$, i.e., the action of the process on the particular subspace,  can still be obtained.
 For all  $| M \kk \in \text{Ker} (A) $, $ \map {R}_{\rho} (M)=A| M \kk=0$. In other words, the process information in $ \text{Ker} (A) $ is inaccessible to us.  Fortunately, one can extract the process information on $  \text{Supp}(A) := \text{Ker}(A)^\perp$ via  the Moore-Penrose pseudoinverse 
\begin{align}
A^\ddag : \,   \text{Ran}(A) \rightarrow  \text{Supp}(A).
\end{align}
 The operator $ A^ \ddag A =: P $ projects states  onto $ \text{Supp}(A) $. Similarly, one can define the pseudoinverse map $\map {R}_{\rho}^\ddag$ via 
\begin{align}
| \map {R}_{\rho} ^\ddag (O ) \kk = A^\ddag |O \kk.
\end{align}
The projection of  $| M \kk $ to $ \text{Supp}(A)  $ can be recovered through
\begin{align}
| \map R_{\rho}^\ddag ( R  )   \kk = A^\ddag  A | M \kk = P |M \kk.
\end{align}
The partially recovered map in general is not completely positive however it contains all the process information on the support of $\rho$.

\subsection{Backward process}
\label{sec:backward process}
 The backward process $\bar M$ can be extracted from $\bar R$ similarly.  Following the same procedure in the previous subsection, we first vectorize the backward PDM $\bar R \, (= \map {R}_{\gamma} (\bar M) )$ in Eq.~\eqref{eq: backwardPDM} and obtain
\begin{align} \label{eq: VectorBackwardPDM}
| \map {R}_{\gamma} ( \bar M)  \kk &= \frac{1}{2} \left( \gamma \otimes \mathbb{1} +  \mathbb{1} \otimes \gamma^T \right)  | \bar M \kk  \nonumber\\
&=:\bar A | \bar M \kk.
\end{align}
If $\gamma $ is full rank, there exists a well-defined inverse map $\map {R}_{\gamma}^{-1}$ and thus $\bar M$ can be extracted via 
$$ | \mathcal{R}_{\gamma}^{-1} (\bar R )  \kk= \bar{A}^{-1}  | \bar R \kk = | \bar M \kk . $$
If $\gamma $ is rank deficient, then we can extract the process information on the support of $\bar A$ via 
$$  | \mathcal{R}_{\gamma}^{\ddag} (\bar R )  \kk= \bar{A}^{\ddag}  | \bar R \kk = \bar {P} | \bar M \kk, $$
where $\bar P = \bar{A}^\ddag   \bar{A}$.

By utilizing  Eq.~\eqref{eq: FBrelation}, a relation between $M$ and $\bar M$ can be found. Vectorizing both sides of Eq.~\eqref{eq: FBrelation}, we have
\begin{align}
\bar A \, | \bar{M} \kk  = & |S R S^\dag \kk =  S \otimes S^* | R \kk  
= (S \otimes S^* )A |M \kk .
\end{align}

\section{Protocol for inferring arrow}
\label{sec:beyongPDM}

Our method for answering Alice's question of inferring the arrow of time in the PDM formalism can be summarised as the following theorem.
\begin{thm}\label{thm: InferArrow}
   If $\rho$ and $\gamma$ are full rank and the extracted $M$  satisfy $M^{T_1} \geq 0$ while $\bar{M}^{T_1} \geq 0$ does not hold, then $R$ is the one consistent with the forwards time flow. For other situations, we may need more information to determine the time direction. 
\end{thm}

Let us illustrate Theorem~\ref{thm: InferArrow} by the following example. Consider a quantum state $\rho_A=(1- a) \ketbra{0}{0} + a \ketbra{+}{+} (0<a<1)$ that undergoes a decohering channel $\map E$ described by the set  of Kraus operators $\{ \ketbra{0}{0},\ketbra{1}{1} \}$. The final state is then given by $\gamma_B= (1- \frac{a}{2}) \ketbra{0}{0} + \frac{a}{2} \ketbra{1}{1}$. Both $\rho_A$ and $\gamma_B$ are full rank, allowing for the extraction of complete information of $M$ and $\bar M$ from both $R$ and $\bar R$, respectively.   Recalling Eq.~\eqref{eq:CJ} and Eq.~\eqref{eq: forwardPDM}, the  CJ matrix $M$ and the PDM $R$ are given by
\begin{align}
M = \left(                 
\begin{array}{cccc}   
      1 & 0 &0  & 0 \\  
      0  &  0 & 0 & 0 \\
      0  &  0 & 0 & 0 \\
      0  &  0 & 0 & 1 
\end{array}
\right), \ \ 
R = \frac{1}{4}\left(                 
\begin{array}{cccc}   
      4-2a & 0 &a  & 0 \\  
      0  &  0 & 0 & a \\
      a  &  0 & 0 & 0 \\
      0  &  a & 0 & 2a 
\end{array}
\right).  \nonumber
\end{align}
Clearly, $M^{T_1}\geq 0$. Next, according to $\bar{R}= S \, R\,S^\dag$ and the vectorization technique proposed above, we arrive at
\begin{align}
\bar{R} = \frac{1}{4}\left(                 
\begin{array}{cccc}   
      4-2a & a &0  & 0 \\  
      a  &  0 & 0 & 0 \\
      0  &  0 & 0 & a \\
      0  &  0 & a & 2a 
\end{array}
\right), \,
\bar{M} = \left(                 
\begin{array}{cccc}   
      1 & \frac{a}{4-2a} &0  & 0 \\  
      \frac{a}{4-2a}  &  0 & 0 & 0 \\
      0  &  0 & 0 & \frac{1}{2} \\
      0  &  0 & \frac{1}{2} & 1 
\end{array}
\right). \nonumber
\end{align}
It is straightforward to see that $\bar{M}^{T_1} \geq 0$ does not hold since here $\bar M = {\bar M}^{T_1} $ and $\bar M$ has two negative eigenvalues  $ \frac{1-\sqrt{2}}{2}, \frac{1}{2}- \frac{\sqrt{a^2+(2-a)^2}}{2(2-a)} .$ Therefore we conclude that the time direction is $A \rightarrow B $ and $R$ is the one consistent with the forwards time flow.

There are two scenarios where additional information is needed to infer the direction of time. Firstly, in situations when both $\rho$ and $\gamma$ are full rank and the two extracted Choi matrices, $M^{T_1} \geq 0$ and $\bar{M}^{T_1} \geq 0$, due to symmetric boundary conditions, the corresponding PDM is consistent with either direction.  Secondly, in cases where either $\rho$ or $\gamma$ is rank deficient, only partial information about the process can be obtained from the corresponding PDM. Consequently, access to $M^{T_1}$ or $\bar{M}^{T_1}$ is restricted, making it challenging for us to infer the arrow of time.
The additional information that may be required to infer the arrow of time is the more fine-grained temporal probabilities when the positivity of the extracted Choi matrices cannot provide an answer. We illustrate this point with the following example, where the boundary condition is symmetric, and the quantum states $\rho, \gamma$ are rank deficient.

Suppose a pure state $\ket{0}$ undergoes the identity channel. $\rho,\gamma =\ket{0}\bra{0}$ and thus are rank deficient. To construct the corresponding 2-time PDM, Pauli operators are measured at times $t_A, t_B$. The constructed forward and backward PDMs are the same, i.e., 
\begin{align}
R_{AB} = \left(                 
\begin{array}{cccc}   
      1 & 0 &0  & 0 \\  
      0  &  0 & 1/2 & 0 \\
      0  &  1/2 & 0 & 0 \\
      0  &  0 & 0 & 0 
\end{array}
\right)= \bar{R}_{AB}.
\end{align}
Distinguishing the arrow of time would be an impossible mission if we were only given the two PDMs, $R_{AB}$ and $\bar{R}_{AB}$.

\begin{figure}
  \centering
  \includegraphics[scale=0.20]{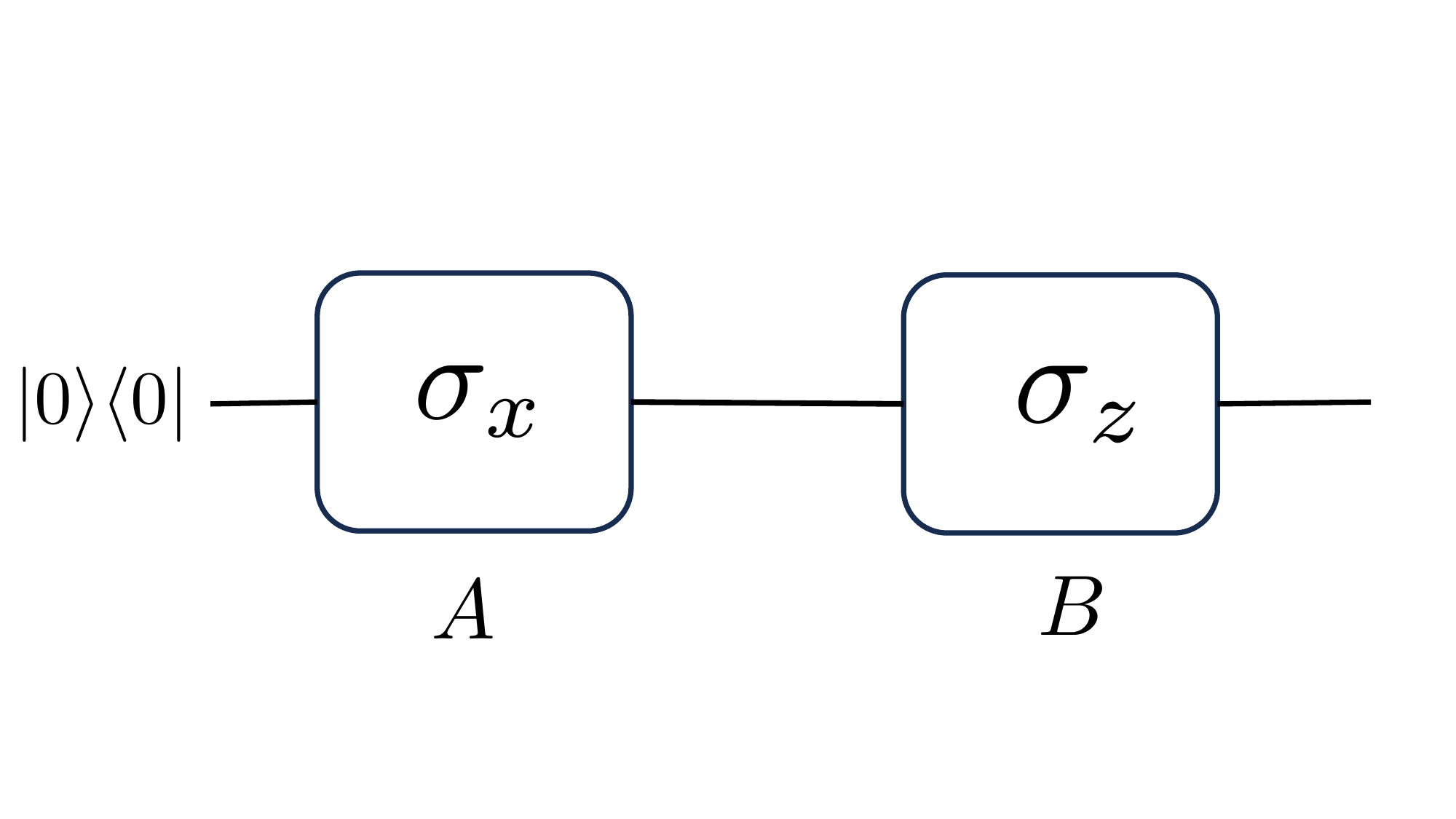}
  \caption{A simple example of measurements whose statistics are sufficient to determine the direction of time. In the actual process the pure state $\ket{0}$ undergoes a $\sigma_x$ measurement at time $A$, an identity channel, and finally a $\sigma_z$ measurement at time $B$. If one is given a table of the measured expectation values where the time order has been flipped so that $B$ is before $A$, one can tell this is not a realizable process.}
  \label{fig:2}
\end{figure}

 Fortunately, there exists a simple way to tell the direction of time. As illustrated in FIG.~\ref{fig:2}, suppose that $\sigma_x$ is measured at $t_A$ and $\sigma_z$ is measured at $t_B$. Denote the expectation of $\sigma_z$ conditioning on $\sigma_x$ and the expectation of $\sigma_x$ conditioning on $\sigma_z$  by $\langle \sigma^B_z \rangle^{\sigma^A_x}$ and  $  \langle \sigma^A_x \rangle^{\sigma^B_z}$, respectively.
When the time direction is $A \rightarrow B$, we have 
\begin{align}
    \langle \sigma^B_z \rangle^{\sigma^A_x}=0= \langle \sigma^A_x \rangle^{\sigma^B_z} .
\end{align}
When the time direction is $B \rightarrow A$, we have 
\begin{align}
    \langle \sigma^B_z \rangle^{\sigma^A_x}=1, \langle \sigma^A_x \rangle^{\sigma^B_z} =0.
\end{align}
In other words, the conditional expectation values aligned with the respective arrows of time differ. This difference gives us the ability to tell the arrow of time. 

Let us go back and examine why the PDM formalism cannot distinguish the arrow of time in some scenarios. To infer the arrow of time, sometimes the fine-grained probabilities, $ P(\Pi^{i_1}_\pm | \Pi^{i_2})$ and $P(\Pi^{i_2}_{\pm} | \Pi^{i_1})$, are needed, as demonstrated above. However, we only have access to the coarse-grained probabilities, $ P(\Pi^{i_1} \Pi^{i_2} =1 ) $ and $P(\Pi^{i_1} \Pi^{i_2} =-1 ) $, from  the correlator $ \langle \{ \sigma_{i_1}, \sigma_{i_2} \} \rangle $ that constructed the PDMs in Eq.~\eqref{eq: DefForwardPDM} via 
\begin{align}
  & P(\Pi^{i_1} \Pi^{i_2} =1 ) -  P(\Pi^{i_1} \Pi^{i_2} =-1 ) =  \langle \{ \sigma_{i_1}, \sigma_{i_2} \} \rangle, \nonumber\\
&  P(\Pi^{i_1} \Pi^{i_2} =1 ) +  P(\Pi^{i_1} \Pi^{i_2} =-1 ) =1,
\end{align}
where $ P(\Pi^{i_1} \Pi^{i_2}=1 ) := P( \Pi^{i_1}_+,  \Pi^{i_2}_+ )+  P( \Pi^{i_1}_{-},  \Pi^{i_2}_{-} )$ and  $P(\Pi^{i_1} \Pi^{i_2} =-1 )$ is defined similarly. Whilst the PDM also contains marginal data $\langle \{ \sigma_{i_1}, \id \} \rangle $ and  $\langle \{  \id, \sigma_{i_2} \} \rangle $ $(\sigma_{i_1} \neq \id \neq \sigma_{i_2} )$, the marginal data are not sufficient to determine a joint fine-grained distribution $P(\Pi^{i_1}, \Pi^{i_2} )$. Such data would be associated with a different experiment than the type used to build a PDM obeying Eq.~\eqref{eq: DefForwardPDM} and Eq.~\eqref{eq: forwardPDM}.

\section{Arrow of time in other quantum spatiotemporal formalisms.}
We now briefly discuss the applicability of our approach to infer the arrow of time to two closely related formalisms: the Leifer-Spekkens framework~\cite{leifer2006quantum,leifer2008quantum,leifer2013towards} and the process matrix formalism~\cite{oreshkov2012quantum,araujo2015witnessing}.

The Leifer-Spekkens framework results from reframing quantum theory as a theory of Bayesian inference and centers around a space-time state defined as
\begin{align}
    L= \sqrt{\rho} \otimes \id_2 \, M \,  \sqrt{\rho} \otimes \id_2,
\end{align}
where $M$ denotes the CJ matrix of a CPTP map $\map E$.
Let $\gamma = \map E (\rho)$. The time reversal of $L_{12}$ is naturally given by
\begin{align}
    \bar{L} =S L S^\dag = \sqrt{\gamma} \otimes \id_1 \, \bar{M} \,  \sqrt{\gamma} \otimes \id_1,
\end{align}
where $\bar M$ denotes the CJ matrix of the time-reversal of $\map E$ in the Leifer-Spekkens framework.
We are unclear on the operational meaning of $L$ and accordingly how it relates to the set-up in FIG~\ref{fig:setup}. The CJ matrices can however be extracted via the same vectorization approach proposed in section~\ref{sec:dynamics}. It turns out that $\bar{M}$ corresponds, not to the CJ matrix of our unitary dilation map, but to the CJ matrix of the Petz recovery map~\cite{petz1986sufficient,petz1988sufficiency}. As a result, the partial transpose $\bar{M}^{T_1}$ is also positive.  Consequently, there cannot be a statement that is very similar to theorem~\ref{thm: InferArrow} for this space-time state. 

The process matrix formalism is an operational framework building on the assumption that causal order is not a fundamental ingredient of nature~\cite{oreshkov2012quantum,araujo2015witnessing}. Given our focus on the arrow of time, we examine the most general bipartite scenario of a definite causal order described by the process matrix formalism. That is a quantum channel with memory, i.e., Alice operates on one part of a correlated state, and her output, along with the other part, is sent to Bob through a channel. This is described by the process matrix of the form,
\begin{align}
    W^{A_1A_2B_1B_2}= W^{A_1A_2B_1} \otimes \id^{B_2},
\end{align}
where $A_1$ and $A_2$ denote Alice's input and output Hilbert spaces, respectively, and similarly for $B_1, B_2$. The corresponding reverse process matrix has the form
\begin{align}
    \bar W= \id^{A_1}\otimes \bar{W}^{A_2B_1B_2}. 
\end{align}
Then, for many cases, the arrow of time can be immediately identified from the process matrix because the output of the later-time party is always the identity operator $\id$ in the forward direction. We leave the analysis of what the precise form of $\bar W^{A_1A_2B_1B_2}$ should be for future studies.

\section{Summary and outlook}

We explored how to establish the time-ordering of measurement data obtained from quantum experiments involving two times and any number of qubits.  We formulated the arrow of time inference problem. We examined conditions on the initial and final states that were symmetric or asymmetric under time reversal. The spatiotemporal measurement data was then represented using a pseudo density matrix. There was a forward process that was CPTP and a reverse process that was obtained via a novel recovery map based on inverting unitary dilations. In cases of asymmetric conditions, the protocol determined whether the data was consistent with the unitary dilation recovery map or the CPTP map. Meanwhile, for symmetric conditions, the recovery map yielded a valid unitary and the experiment may have taken place in either direction. More data was needed for solving the arrow inference problem in the unitary case as well as in the case of states that are not full rank. We also considered the applicability of this approach to the Leifer-Spekkens or Process matrix space-time states.

Apart from the relation to other space-time states discussed above, another question that emerges is: Can the positive unitary dilation recovery map that appeared naturally in this set-up perform better than the Petz recovery map in quantum information tasks such as Bayesian quantum parameter estimation? Secondly, the calculations revealed an unexpected consequence that, whilst the normal quantum density matrix can be fully characterised by coarse-grained measurements giving correlations between Pauli matrices, such coarse-grained data concerning temporally separated events cannot be used to fully determine the probability distributions over the temporally separated outcome in an analogous manner,  suggesting a further distinction between the spatial and temporal directions which deserves investigation. Thirdly,
it would be interesting to quantify the arrow of time in the PDM $R$, and one possible approach is through the following measure
\begin{align}
\map{A}(R)= F( \bar{M}^{T_1}_{\bar R} ) - F(M^{T_1}_{R}),
\end{align}
where $F(O) :=\text{Tr} (\sqrt{OO^\dag }-O)$. Under this measure,
 $\map{A}(R) >0 $ means that $R$ is the forward PDM;
       $\map{A}(R) <0 $ means that $\bar R$ is the forward PDM whereas
     $\map{A}(R) =0$ means that the arrow of time could be in either direction. General properties of the measure $\map A$ or other possible measures of the arrow of time should be investigated further. For example, $\map {A}(R) $ or a variant thereof may constitute a natural measure of thermodynamical irreversibility.

\section*{Acknowledgements} We thank Caslav Brukner, Zhenhuan Liu, Zhian Jia, and Fei Meng for the discussions. We also thank Masahito Hayashi for his feedback. XL and OD acknowledge support from the National Natural Science Foundation of China (Grants No.12050410246, No.1200509, No.12050410245) and City University of Hong Kong (Project No. 9610623). QC acknowledges support from the QuantERA II Programme that has received funding from the European Union’s Horizon 2020 research and innovation programme under Grant Agreement No 101017733.

\bibliography{ref}

\end{document}